\documentclass[a4paper,10pt]{article}
\usepackage[utf8]{inputenc}

\title{Holographic  Cosmology from BIonic Solutions
}
\author{Alireza Sepehri $^{1,2}$ , Mir Faizal $^{3}$ , \\
Mohammad Reza Setare $^{4}$ ,  Ahmed Farag Ali$^{5}$
 \\\\$^1$ Faculty of Physics, Shahid Bahonar University, \\P.O. Box 76175, Kerman, Iran.\\$^{2}$ Research Institute for Astronomy and Astrophysics of
 Maragha (RIAAM),\\ P.O. Box 55134-441, Maragha, Iran.\\$^{3 }$
Department of Physics and Astronomy, University of Waterloo,\\ Waterloo, ON, N2L 3G1, Canada.
 \\ $^4$  Department of Science, Campus of Bijar,\\
University of Kurdistan, Bijar, Iran.\\ $^5$  Deptartment of Physics, Faculty of Science, \\ Benha University, Benha 13518, Egypt.
}\date{}
\begin{document}

\maketitle
\begin{abstract}
In this paper, we will use   a   BIonic solution for analysing
the holographic  cosmology.
A BIonic solution is a configuration of a D3-brane and an  anti-D3-brane connected by a wormhole,
and holographic cosmology is a recent proposal to explain cosmic expansion by using
the holographic principle.
In our model,   a  BIonic configuration will be produced  by the transition of  fundamental black strings.
The formation of a BIonic configuration will cause  inflation.
As the
  D3-brane moves away from the anti-D3-brane,
the  wormhole will get annihilated, and the  inflation will end with the annihilation of this wormhole.
However, it is possible for a D3-brane to collide with an anti-D3-brane.
Such a collision will occur if the   distance between the D3-brane and the anti-D3-brane  reduces,
and this will create  tachyonic states. We will demonstrate that these
tachyonic states will lead to the formation of a  new wormhole, and this
 will  cause acceleration of the universe before such a collision.
\end{abstract}

 \maketitle
\section{Introduction}
The low energy effective field theory action for D-branes is the Dirac-Born-Infeld (DBI) action
 \cite{a}-\cite{a1}.
This non-linear action can be used to construct a BIonic solution.
A BIonic solution is
a configuration of a D3-brane and an  anti-D3-brane connected by a wormhole \cite{b}-\cite{c}. In a BIonic solution,
the F-string end on a point  of a D-brane, and  the F-string charge gets associated with the world-volume electric flux carried by the D-brane.
  Thus, the  F-strings (which are one-dimensional objects) become  higher-dimensional
brane wrapped on a sphere. This phenomena   critically depends on the non-linearity of the DBI action. Such brane configurations has also been used for
analysing various aspects of the
 $AdS/CFT$ correspondence \cite{d}-\cite{f}, including
    giant gravitons \cite{g}-\cite{h}. This is because it is possible for
 gravitons (which satisfy the BPS bound)   to become giant gravitons, if they
 are moved on the equator of the five sphere in the $AdS_5 \times S^5$ background \cite{g}-\cite{h}.
 It may be noted that  Wilsons loops have also been used for  analysed   various D-branes configurations \cite{i}-\cite{j}.
 Thus, D-brane configurations have been used for analyzing various interesting physical systems.
 In this paper, we will use the BIonic solutions  for analysing the holographic cosmology.

There is a close relation between geometry and thermodynamics, and this
relation between the geometry and thermodynamics is the basis of the   Jacobson's  formalism \cite{4}.
In this formalism, gravity is described as an emergent thermodynamic force, and
  the Einstein equation are obtained from the
Clausius relation.  Thus, the structure of spacetime become an emergent
structure in this formalism. It may be noted that
 this thermodynamic approach   has led to the development of the Verlinde formalism \cite{8}. Furthermore, the   holographic cosmology is based on this
  formalism and the holographic principle \cite{1}-\cite{2}. The holographic principle states   that the number of degrees of freedom
in a region of space is equal to the number
 of degrees of freedom on the boundary surrounding that region of space. So,  it is possible to
 use the     difference
 between the    degrees of freedom in a region of space and the degrees of freedom on the
 boundary surrounding that region of space to explain the cosmic acceleration, and this  proposal is called the holographic cosmology \cite{1}-\cite{2}.

The original proposal of holographic cosmology   has been used in several modified theories of gravity. In fact, the
  Friedmann equations in Gauss-Bonnet gravity (and even more general Lovelock gravity) have been obtained
using  the holographic  cosmology  \cite{9}. Furthermore, the brane world models \cite{b0}, cosmological models in scalar-tensor gravity \cite{b1},
and cosmological models in $f(R)$ gravity \cite{b2}, have been studied using the Jacobson's thermodynamic approach.
As the holographic cosmology is  based on the Jacobson's thermodynamic
approach,   brane world models, cosmological model in scalar tensor gravity, and  cosmological model in  $f(R)$ gravity,
 have been analysed using the holographic  cosmology \cite{f1}. A generalization of the original
 proposal for the holographic cosmology has been used for deriving
 Friedmann equation corresponding to the Friedmann-Robertson-Waker universe with an arbitrary spatial curvature \cite{f2}.
The Friedmann equation  for Gauss-Bonnet gravity (and even more general Lovelock gravity) with an arbitrary spatial curvature has
   also been derived using this generalization proposal \cite{f4}.
 However, it has been demonstrated that such a generalization
 is only valid if the aerial volume is used instead of the proper volume \cite{f6}.

 It has also been possible to   analyse the holographic cosmology using the BIonic solution \cite{f8}.
 In this model, the  D3-brane represent   the universe,
 and the degrees of freedom on this brane are controlled by the evolution of BIonic solution.
It may be noted that if a  D3-brane is away from  an anti-D3-brane, then the spike of the  D3-brane are separated from the
 spike of the   anti-D3-brane. However, as the distance between the D3-brane and the anti-D3-brane reduces to a critical value,
  the spike of D3-brane meets the spike of the anti-D3-brane, and a wormhole is formed  \cite{f9}-\cite{ff10}.
  This configuration of a D3-brane and an anti-D3-brane joined by a wormhole is called a BIon. It is possible for this
  wormhole to act as a channel for the degrees of freedom to flow into the D3-brane, and this can cause the cosmic acceleration.
So,  we will analyse the holographic cosmology using
   a BIonic solution \cite{f9}-\cite{f10}. In fact, we will use a thermal generalization of the usual BIonic solution for
   analyzing holographic cosmology. This is because the BIon is a static solution, but we need a dynamic parameter in this solution
   to relate it to the time evolution of our universe.  As the    temperature of the universe decreases during its  evolution,
   we can identify the cosmological clock with the temperature of the BIonic solution.
In our model, the BIon will first form
 from  fundamental black strings.
 The D3-brane in this BIon will represent our universe. The inflation will occur because the
 degrees of freedom  will flow into the  the D3-brane. As the    D3-branes moves away from  the anti-D3-brane,
  the spike
 of  D3-brane will get separating from the  spike of anti-D3-brane, and this will annihilate  the wormhole. The inflation will end when the  wormhole
 gets annihilated.
   However,  it is possible for a D3-brane to collide with an anti-D3-brane. We will demonstrate that tachyonic states will form as
 a D3-brane approaches an anti-D3-brane,
 and the existence of these tachyonic states will lead to the formation
 of a new wormhole. This will cause the late time   acceleration of the universe before such an collision.

The   the paper is organized as the follows.  In section
\ref{o1}, we discuss the emergence of BIon from  fundamental black strings,
and use it for analyzing holographic inflationary cosmology.
In
section \ref{o2}, we analyse  the tachyonic states in theory and analyse their consiquences.
In the last section, we summarize our results and discuss some possible extensions of the results obtained in this paper.

\section{Holographic BIonic Solutions }\label{o1}
In this section, we will analyse the holographic cosmology using BIonic solutions.
We will   start from  $k$ fundamental
black strings, and    a BIon will be formed from a   transition of these $k$ fundamental
black strings.  The
 supergravity
solution for $k$ coincident non-extremal black F-strings (lying
along the $z$ direction) can be written
\begin{eqnarray}
 ds^{2} &=& H^{-1}(-f dt^{2} + dz^{2})+ f^{-1}dr^{2} + r^{2}d\Omega_{7}^{2},\nonumber\\
  e^{2\phi} &=&H^{-1},\: B_{0} = H^{-1}-1,\nonumber\\
  H &=& 1 +
\frac{r_{0}^{6}\sinh^{2}\alpha}{r^{6}},\nonumber \\ f&=&1-\frac{r_{0}^{6}}{r^{6}}.
\label{m1}
\end{eqnarray}
 The mass density (along
the $z$ direction) for this solution is given by   \cite{f10},
\begin{eqnarray}
&& \frac{dM_{F1}}{dz} = T_{F1}k +
\frac{16(T_{F1}k\pi)^{3/2}T^{3}}{81T_{D3}}+
\frac{40T_{F1}^{2}k^{2}\pi^{3}T^{6}}{729T_{D3}^{2}}.\label{m2}
\end{eqnarray}
It may be noted that this equation is  valid only up to order $T^6$ (using small temperature limit).
We will use this approximation through out this paper, and higher order corrections will
produce corrections terms to the expressions obtained in this paper. However, they will not change
any quantitative feature of the our model.
The metric in which the  BIon  is embedded can be written as
\cite{f9},
\begin{eqnarray}
&& ds^{2} = -dt^{2} + dr^{2} + r^{2}(d\theta^{2} + \sin^{2}\theta
d\phi^{2}) + \sum_{i=1}^{6}dx_{i}^{2}. \label{m3}
\end{eqnarray}
 Choosing the worldvolume coordinates of the D3-brane as
$\lbrace\sigma^{a}, a=0..3\rbrace$ and defining $\tau =
\sigma^{0},\,\sigma=\sigma^{1}$, the coordinates of BIon are given
by \cite{f9},
\begin{eqnarray}
t(\sigma^{a}) =
\tau,\,r(\sigma^{a})=\sigma,\,x_{1}(\sigma^{a})=z(\sigma),\,\theta(\sigma^{a})=\sigma^{2},\,\phi(\sigma^{a})=\sigma^{3}.
\label{m4}
\end{eqnarray}
The remaining coordinates $x_{i=2,..6}$ are constant in our model. The
embedding function $z(\sigma)$  describes the bending of the
brane. The induced metric on the
brane can then be written as
\begin{eqnarray}
\gamma_{ab}d\sigma^{a}d\sigma^{b} = -d\tau^{2} + (1 +
z'(\sigma)^{2})d\sigma^{2} + \sigma^{2}(d\theta^{2} +
\sin^{2}\theta d\phi^{2}) \label{m5},
\end{eqnarray}
where $z$ is a transverse coordinate to the branes and $\sigma$
is its radius. So, the
  spatial volume element can be expressed as  $dV_{3}=\sqrt{1 +
z'(\sigma)^{2}}\sigma^{2}d\Omega_{2}$. Now we  impose the  boundary
conditions,  $z(\sigma)\rightarrow 0$ for $\sigma\rightarrow
\infty$ and $z'(\sigma)\rightarrow -\infty$ for $\sigma\rightarrow
\sigma_{0}$, where  $\sigma_{0}$ is the minimal two-sphere radius.
 The mass density of the BIon along the
$z$ direction is given by
\begin{eqnarray}
 &&  \frac{dM_{BIon}}{dz} = T_{F1}k + \frac{3\pi T_{F1}^{2}k^{2}
T^{4}}{32T_{D3}^{2}\sigma_{0}^{2}}+
 \frac{7\pi^{2} T_{F1}^{3}k^{3} T^{8}}{512T_{D3}^{4}\sigma_{0}^{4}} \label{m6}.
\end{eqnarray}
This equation is  valid only up to order $T^6$, and thus all the equations obtained using this equation
are also valid to the order $T^6$.
It may be noted that  the  mass density  of the  BIon given   is equal to     the mass
density   of the F-strings  at $\sigma = \sigma_{0}$.  This is because the
thermal BIon  is formed from   $k$ F-strings at $\sigma =
\sigma_{0}$.
Thus, at this point we can identify the thermodynamics of the
BIon with the thermodynamics of $k$ F-strings.  To match two
Eq.  (\ref{m2}) with Eq.  (\ref{m6}) at this point,
$\sigma_{0}$ should have the following temperature dependence
\begin{eqnarray}
&& \sigma_{0} =
\left(\frac{\sqrt{kT_{F1}}}{T_{D3}}\right)^{1/2}\sqrt{T}\left[C_{0} +
C_{1}\frac{\sqrt{kT_{F1}}}{T_{D3}}T^{3}\right]\label{m7},
\end{eqnarray}
where $T_{F1} = 4k\pi^{2}T_{D3}g_{s}l_{s}^{2}$, and  $C_{0}$, $C_{1}$,
$F_{0}$, $F_{1}, F_{2}$ are numerical coefficients which can
be obtained  from   $T^{3}$ and $T^{6}$ terms in
Eqs. (\ref{m2}) and (\ref{m6}).     It may be noted that from
Eq. (\ref{m7}),  $\sigma$   is  a function of  the temperature, and so the metric is a dynamic function of the temperature.
Furthermore, $\sigma$ can also be related to the width of the wormhole, and   so
  at a critical temperature  $T_{end}$,
   the wormhole gets   annihilated
\begin{eqnarray}
&& \sigma_{0}=0,
\bar{C}_{0}=-C_{0}\rightarrow T_{end}=\frac{\bar{C}_{0}\sqrt{T_{D3}}}{C_{1}kT_{F1}}.
\label{m10}
\end{eqnarray}
It may be noted that we have identified the temperature of the BIonic solution with the cosmological clock. This is because the temperature of the
BIonic solution is very high at the beginning  of inflation, and then reduces with the evolution of the universe. Hence, a cosmological
clock can be identified with the temperature of the BIonic solution \cite{f9}-\cite{ff10}.
In fact, the  temperature is infinite at the
beginning, and it    decreases    to $T_{end}$
at the end of inflation. The  width of the   wormhole also decreases with time, and at $T=T_{end}$ the
the wormhole is annihilated.
The annihilation of the wormhole occurs due to the D3-brane moving away from the anti-D3-brane. This is because when the  D3-brane is close to  the anti-D3-brane there
   spikes meet, forming a wormhole. However, as the D3-brane moves away from the anti-D3-brane, there spikes move away from each other. So,
     at a critical point the wormhole gets annihilated.

The  inflation occurs because the wormhole acted as a channel for the degrees of freedom  to flow into the
 D3-brane.
 Putting $k$   F-string
charges along the radial direction and using Eq. (\ref{m7}),
we obtain \cite{f10},
\begin{eqnarray}
z(\sigma)= \int_{\sigma}^{\infty}
d\acute{\sigma}\left(\frac{F(\acute{\sigma})^{2}}{F(\sigma_{0})^{2}}-1\right)^{-\frac{1}{2}}.
\label{m9}
\end{eqnarray}
 Now for a thermal BIon,  $F(\sigma)$ can be expressed as
\begin{eqnarray}
F(\sigma) = \sigma^{2}\frac{4 \cosh^{2}\alpha - 3}{ \cosh^{4}\alpha},
\label{m10}
\end{eqnarray}
where $ \cosh\alpha$ is given by
\begin{eqnarray}
 \cosh^{2}\alpha = \frac{3}{2}\frac{\cos\frac{\delta}{3} +
\sqrt{3}\sin\frac{\delta}{3}}{\cos\delta} \label{m11},
\end{eqnarray}
Here we have defined $\cos\delta  $ as
\begin{eqnarray}
\cos\delta   \equiv  \overline{T}^{4}\sqrt{1 +
\frac{k^{2}}{\sigma^{4}}},&&
\overline{T}  \equiv
\left(\frac{9\pi^{2}N}{4\sqrt{3}T_{D_{3}}}\right)T, \nonumber \\ k \equiv \frac{k
T_{F1}}{4\pi T_{D_{3}}} \label{m12}. &&
\end{eqnarray}
where  $T$ is the   temperature of BIon, $N$ is the
number of D3-branes,  $T_{D_{3}}$ is the tension of the brane,  and $T_{F1}$ is  tensions of
the fundamental string. It is possible to attach  a mirror
solution to Eq. (\ref{m9}), and  obtain a  wormhole configuration.
Now $\bar{\Delta} = 2z(\sigma_{0})$, is the    distance  separating
 $N$ D3-branes from $N$ anti-D3-branes for a given BIonic configuration.
 This configuration is thus defined by   four parameters $N$, $k$, $T$ and
$\sigma_{0}$,
\begin{eqnarray}
\bar{\Delta} = 2z(\sigma_{0})= 2\int_{\sigma_{0}}^{\infty}
d\acute{\sigma}\left(\frac{F(\acute{\sigma})^{2}}{F(\sigma_{0})^{2}}-1\right)^{-\frac{1}{2}}.
\label{m13}
\end{eqnarray}
In   the   small temperature limit, we obtain
\begin{eqnarray}
\bar{\Delta} =
\frac{2\sqrt{\pi}\Gamma(5/4)}{\Gamma(3/4)}\sigma_{0}\left(1 +
\frac{8}{27}\frac{k^{2}}{\sigma_{0}^{4}}\overline{T}^{8}\right).
\label{m14}
\end{eqnarray}

Now we can construct a holographic cosmological model using  this BIonic solution    \cite{1, 2, f8}.
The total entropy during cosmic expansion  can be obtained by adding the
number of degrees of freedom in the bulk  with
 the number of degrees of freedom on the boundary.
It may be noted that by surface degrees of freedom we
mean the degrees of freedom on the apparent horizon of the universe.
 Furthermore, according to the holographic cosmology,  the difference between the
 number of degrees of freedom in the bulk and the number of degrees of freedom on the boundary is equal to
 the  mass density \cite{1, 2, f8}. We can obtain the mass density by placing  the wormhole
  along the $z$-axis \cite{1, 2, f8}. In
  this paper, this sum is equated with the total
 degrees of freedom of the BIon, as we are using a BIon to model the inflation.
 Thus, we can write the following equations,
 \begin{eqnarray}
 N_{sur} + N_{bulk} &=& N_{BIon}= N_{brane} + N_{anti-brane} +
N_{wormhole}\nonumber \\& \simeq&
4l_{P}^{2}S_{BIon}\nonumber \\ &=& \frac{4T_{D3}^{2}}{\pi T^{4}}\int
d\sigma\frac{F(\sigma)}{\sqrt{F^{2}(\sigma)-F^{2}(\sigma_{0})}}\sigma^{2}\frac{4}{ \cosh^{4}\alpha}
\nonumber \\ N_{sur} - N_{bulk} &\simeq& \int d\sigma
\frac{dM_{BIon}}{dz}\nonumber \\ &=& \frac{2T_{D3}^{2}}{\pi T^{4}}\int d\sigma
\frac{F(\sigma)}{F(\sigma_{0})}\sigma^{2}\frac{4 \cosh^{2}\alpha +
1}{ \cosh^{4}\alpha}. \label{m15}
\end{eqnarray}
According to this model, our universe represented by a D3-brane
interacts with an anti-D3-brane through a wormhole. Thus, the total
number of degrees of freedom is equal to sum of the  number of degrees of
freedom of the D3-brane, the anti-D3-brane, and  the wormhole,
$N_{total}=N_{sur,universe}+N_{sur,anti-universe}+N_{wormhole}$.
Here the anti-D3-brane represents the anti-universe.
So, we can define
$N_{bulk}=N_{sur,anti-universe}+N_{wormhole}$. Now as our universe is
located on a brane and another anti-universe is located  on an
anti-brane, we can write
$N_{total}=N_{sur,universe}+N_{bulk}=N_{sur,universe}+N_{sur,anti-universe}+N_{wormhole}=N_{brane}
+ N_{anti-brane} + N_{wormhole}$ . Solving these equations
simultaneously, we obtain,
\begin{eqnarray}
 N_{sur} &\simeq& \frac{4T_{D3}^{2}}{\pi T^{4}}\int
d\sigma\frac{F(\sigma)}{\sqrt{F^{2}(\sigma)-F^{2}(\sigma_{0})}}\sigma^{2}\frac{4}{ \cosh^{4}\alpha}
\nonumber \\&& + \frac{2T_{D3}^{2}}{\pi T^{4}}\int d\sigma
\frac{F(\sigma)}{F(\sigma_{0})}\sigma^{2}\frac{4 \cosh^{2}\alpha +
1}{ \cosh^{4}\alpha}, \nonumber \\
N_{bulk}& \simeq&
\frac{4T_{D3}^{2}}{\pi T^{4}}\int
d\sigma\frac{F(\sigma)}{\sqrt{F^{2}(\sigma)-F^{2}(\sigma_{0})}}\sigma^{2}\frac{4}{ \cosh^{4}\alpha}
\nonumber \\&& -\frac{2T_{D3}^{2}}{\pi T^{4}}\int d\sigma
\frac{F(\sigma)}{F(\sigma_{0})}\sigma^{2}\frac{4 \cosh^{2}\alpha +
1}{ \cosh^{4}\alpha} \label{m16}.
\end{eqnarray}

We can obtain
 cosmological parameters like Hubble parameter and energy density of the universe from this  BIonic solution.
  According to the usual holographic cosmology, the number of
degrees of freedom on the spherical surface of apparent horizon
with radius $r_{A}$ is proportional to its area \cite{f2},
\begin{eqnarray}
&& N_{sur} = \frac{4\pi r_{A}^{2}}{l_{P}^{2}}\label{m17},
\end{eqnarray}
where $r_{A} = 1/ \sqrt{H^{2} + \frac{\bar{k}}{a^{2}}}$ is  the radius of the
apparent horizon  for the Friedmann-Robertson-Waker universe, and $H=
\frac{\dot{a}}{a}$ is the Hubble parameter ($a$ is the scale
factor). So,  we   obtain an expression for     the Hubble parameter,
\begin{eqnarray}
&& H_{flat,inf} \simeq \frac{18\pi k^{4}N^{12}
T_{F1}^{14}}{T_{D3}^{14}\sigma_{0}^{8}} T^{12} + \frac{8\pi
k^{2}N^{10} T_{F1}^{12}}{T_{D3}^{12}\sigma_{0}^{4}}.
T^{10}\label{m18}.
\end{eqnarray}
The temperature of the universe  is very high at its
beginning, and   the rate of expansion is also very large.
This can be seen from the fact the at this stage there is a
large Hubble parameter. This is because the wormhole
acts as a channel for the degrees of freedom to flow into the  universe.
However, at a critical temperature, the inflation ends because the wormhole get annihilated at that
critical temperature.

We can calculate the energy density
of the universe using the Friedmann equation for the flat Friedmann-Robertson-Waker universe,
\begin{eqnarray}
&& \rho_{flat,inf} = \frac{3}{8\pi l_{P}^{2}}H_{flat}^{2} \simeq \frac{27\pi
k^{8}N^{24} T_{F1}^{28}}{4l_{P}^{2}T_{D3}^{28}\sigma_{0}^{16}}
T^{24} + \frac{3\pi k^{4}N^{20}
T_{F1}^{24}}{l_{P}^{2}T_{D3}^{24}\sigma_{0}^{8}}
T^{20}\label{m19}.
\end{eqnarray}
Thus,  the energy density depends on
the temperature of the BIon. So, the energy density of the universe reduces as
the temperature of the BIon reduces to $T=T_{end}$, at the end of  the  inflation.
Now using (\ref{m17}),
and assuming $r_{A} = 1/ \sqrt{H^{2} + \frac{\bar{k}}{a^{2}}}$, we can  obtain Hubble parameter
for non-flat universe in terms of temperature,
\begin{eqnarray}
&& H_{o/c,inf} \simeq \sqrt{\left(\frac{18\pi k^{4}N^{12}
T_{F1}^{14}}{T_{D3}^{14}\sigma_{0}^{8}} T^{12} + \frac{8\pi
k^{2}N^{10} T_{F1}^{12}}{T_{D3}^{12}\sigma_{0}^{4}}
T^{10}\right)^{2}-\bar{k}/a^{2}}\label{mm19}.
\end{eqnarray}
Furthermore, we define $ \mathcal{T}$ as
\begin{eqnarray}
 \mathcal{T} = \left(\frac{18\pi k^{4}N^{12}
T_{F1}^{14}}{T_{D3}^{14}\sigma_{0}^{8}} T^{12} + \frac{8\pi
k^{2}N^{10} T_{F1}^{12}}{T_{D3}^{12}\sigma_{0}^{4}}
T^{10}\right)^{2},
\end{eqnarray}
and solve this equation. Now  we can write  the scale factor for open $(k=-1)$ universe as
\begin{eqnarray}
&& a_{o,inf}(t) \simeq \exp -\int dt [ \mathcal{T}  + ln(t)  ]\label{mmm19},
\end{eqnarray}
  and the scale factor for closed $(k=+1)$ universe as
\begin{eqnarray}
&& a_{c,inf}(t) \simeq \exp -i\int dt [  \mathcal{T} + ln(t) + \frac{\pi}{2}  ]\label{mmmm19}.
\end{eqnarray}
 Thus, the  behavior of open and closed universes also depends on  the temperature.
The scale factor of open universe is almost zero at the beginning ($T=\infty$),  and it increases with decreasing
 temperature. At the end of the inflation, it has a larger value. However,  the scale factor of closed
 universe oscillates during inflation.
 The energy density for open and closed universes during this inflation can be written as
 \begin{eqnarray}
  \rho_{o/c,inf} &=& \frac{3}{8\pi l_{P}^{2}}(H_{o/c}^{2}+k/a^{2})  \nonumber \\&\simeq&
   \frac{3}{8\pi l_{P}^{2}}H_{flat}^{2} \nonumber \\ &\simeq& \frac{3}{8\pi l_{P}^{2}}\left(\frac{27\pi
k^{8}N^{24} T_{F1}^{28}}{4l_{P}^{2}T_{D3}^{28}\sigma_{0}^{16}}
T^{24} + \frac{3\pi k^{4}N^{20}
T_{F1}^{24}}{l_{P}^{2}T_{D3}^{24}\sigma_{0}^{8}}\right)
T^{20}\nonumber \\ &=&\rho_{flat,inf}.
\label{mmmmm19}
\end{eqnarray}
We   now observe that in this model the energy density is the same for       flat, open and closed universes.
This is because that the energy density originates due to the  evolution of a  BIonic solution,
and so it does not depend on type of universe.

It may be noted that when the inflation ends, the  wormhole is annihilated,  and
the mass distribution along $z$-direction is absent at this stage. Thus, at this stage the surface degrees of freedom
  on the boundary and the bulk degrees
of freedom in the universe  can be expressed as
\begin{eqnarray}
&& N_{sur} - N_{bulk} \simeq \int_{\sigma_{0}}^{\sigma_{0}}
d\sigma \frac{dM_{BIon}}{dz}=0.
\end{eqnarray}
So, we can write
\begin{eqnarray}
 N_{sur} &=& N_{bulk} \label{m20}.
\end{eqnarray}
Thus, the  degrees of freedom on
the   boundary is equal to the degrees of freedom in bulk,  at the end of the inflation.
In fact, the inflation was caused because of the difference between the degrees of freedom on the boundary and in the bulk,
and so the inflation ends when the bulk degrees of freedom equals the boundary degrees of freedom.

\section{Tachyonic States}\label{o2}
 In
this section, we will analyse the effect of tachyonic states in
this theory.
It is possible for a D3-brane to collide with an anti-D3-brane. This will occur
when the distance between a D3-brane and an  anti-D3-brane
decreases. However, as the distance between a D3-brane and an anti-D3-brane reduces,     tachyonic states  will be created.
These tachyonic states will form  will      tachyonic spikes, and when the tachyonic spikes of the D3-brane meet
the  tachyonic spikes of the anti-D3-brane, a   tachyonic wormhole will be formed \cite{ff10}. So, this  new BIonic
solution will consist of the tachyonic
wormhole,  the    D3-brane and the anti-D3-brane.
The wormhole in this new BIonic solution will again act as a channel    for  the
degrees of freedom  to flow  into the universe. This will cause the number
of degrees of freedom in the universe  to increase, and that will  in turn cause late time acceleration of the universe. Thus,
the universe will evolved from non-phantom phase to phantom one.
Now we consider a set of
D3-branes and anti-D3-branes   (\ref{m3}),  which
are placed at points $z_{1} = l/2$ and $z_{2} = -l/2$,  respectively.
For this system, we can write
 \begin{eqnarray}
 S&=&-\tau_{3}\int d^{9}\sigma \sum_{i=1}^{2}
V(TA,l)e^{-\phi}(\sqrt{-det A_{i}}),\nonumber \\
(A_{i})_{ab}&=&(g_{MN}-\frac{TA^{2}l^{2}}{Q}g_{Mz}g_{zN})\partial_{a}x^{M}_{i}\partial_{b}x^{M}_{i}
\nonumber \\ && +F^{i}_{ab}+\frac{1}{2Q}((D_{a}TA)(D_{b}TA)^{\ast}+(D_{a}TA)^{\ast}(D_{b}TA))\nonumber
\\&&
+il(g_{az}+\partial_{a}z_{i}g_{zz})(TA(D_{b}TA)^{\ast}\nonumber \\ && -TA^{\ast}(D_{b}TA))+
il(TA(D_{a}TA)^{\ast}-TA^{\ast}(D_{a}TA))\nonumber
\\&&\times (g_{bz}+\partial_{b}z_{i}g_{zz})\left(1-\frac{\pi^{2}N
T^{4}}{6T_{D3}}\right). \label{m21}
\end{eqnarray}
where
  \begin{eqnarray}
&& Q=1+TA^{2}l^{2}g_{zz}, \nonumber \\&&
D_{a}TA=\partial_{a}TA-i(A_{2,a}-A_{1,a})TA,
V(TA,l)=g_{s}V(TA)\sqrt{Q}, \nonumber \\&& e^{\phi}=g_{s}\left( 1 +
\frac{R^{4}}{z^{4}} \right)^{-\frac{1}{2}} . \label{m22}
\end{eqnarray}
Here  $\phi$ is the dilaton field,  $A_{2,a}$ is the gauge field,  and $F^{i}_{ab}$
is the field strengths on the world-volume of the non-BPS
brane. Furthermore,  $TA$ is the tachyon field, $\tau_{3}$ is the brane tension
and $V (TA)$ is the tachyon potential. The indices $a,b$ denote the
tangent directions of D3-branes, while the indices $M,N$ run over the
background ten-dimensional space-time directions. The D3-brane and
the anti-D3-brane are labeled by $i = 1, 2$,  respectively.
The separation between the  D3-brane and the anti-D3-brane  is denoted by $z_{2} - z_{1}
= l$. Here  we have chosen,
$2\pi\acute{\alpha}=1$. Now we can write a   potential for this system as
\cite{m1}-\cite{m2},
 \begin{eqnarray}
V(TA)=\frac{\tau_{3}}{ \cosh\sqrt{\pi}TA} \label{m23}.
\end{eqnarray}

 Let us only consider the  $\sigma$
dependence of the tachyon field $TA$, and   for simplicity, we neglect the contributions coming from the
gauge fields. So,  the Lagrangian given by Eq. (\ref{m27}), in the
region    $r> R$ and $TA'\sim constant$,  simplifies to
   \begin{eqnarray}
L \simeq-\frac{\tau_{3}}{g_{s}} \int d\sigma \sigma^{2}
V(TA)(\sqrt{D_{1,TA}}+\sqrt{D_{2,TA}})\left(1-\frac{\pi^{2}N
T^{4}}{6T_{D3}}\right) \label{m24},
\end{eqnarray}
where
 \begin{eqnarray}
D_{1,TA} = D_{2,TA}\equiv D_{TA} = 1 + \frac{l'(\sigma)^{2}}{4}+
\dot{TA}^{2} -  TA'^{2}.\label{m25}
\end{eqnarray}
Here we  have  assume that $TA l\ll TA'$. Now we will obtain the  Hamiltonian
corresponding to this Lagrangian,  and used it for analysing this system.
The canonical momentum density $\Pi =
\frac{\partial L}{\partial \dot{TA}}$ associated with the tachyon is given by
 \begin{eqnarray}
\Pi = \frac{V(TA)\dot{TA}}{ \sqrt{1 + \frac{l'(\sigma)^{2}}{4}+
\dot{TA}^{2} -  TA'^{2}}}\left(1-\frac{\pi^{2}N
T^{4}}{6T_{D3}}\right)\label{m26}.
\end{eqnarray}
So,   the Hamiltonian can be written as
\begin{eqnarray}
H_{DBI} = 4\pi\int d\sigma  \sigma^{2} \Pi \dot{TA} - L.
 \label{m27}.
\end{eqnarray}
Thus, by  choosing $\dot{TA} = 2 TA'$, we obtain
\begin{eqnarray}
H_{DBI} = 4\pi\int d\sigma \sigma^{2} [\Pi
(\dot{TA}-\frac{1}{2}TA')] + \frac{1}{2}TA\partial_{\sigma}(\Pi
\sigma^{2}) - L.
 \label{m28}
\end{eqnarray}
Here we have   integrated by parts
the term proportional to $\dot{TA}$. So,  tachyon can
be studied as a Lagrange multiplier by imposing the constraint
$\partial_{\sigma}(\Pi \sigma^{2}V(TA))=0$ on the canonical
momentum. Solving this equation, we obtain
\begin{eqnarray}
\Pi =\frac{\beta}{4\pi \sigma^{2}},
 \label{m29}
\end{eqnarray}
where $\beta$ is a constant.  Now  substituting Eq. (\ref{m35}) in Eq. (\ref{m34}), we
obtain
\begin{eqnarray}
&& H_{DBI} = \int d\sigma V(TA)\left(\sqrt{1 + \frac{l'(\sigma)^{2}}{4}
+ \dot{TA}^{2} -  TA'^{2}}\right)F_{DBI} ,  \nonumber \\&&
F_{DBI}=\sigma^{2}\sqrt{1 +
\frac{\beta}{\sigma^{2}}}\left(1-\frac{\pi^{2}N
T^{4}}{6T_{D3}}\right)\label{m30}.
\end{eqnarray}
The resulting equation of motion for $l(\sigma)$ is given by
\begin{eqnarray}
&&\left(\frac{l'F_{DBI}}{4\sqrt{1+
\frac{l'(\sigma)^{2}}{4}}}\right)'=0\label{m31}.
\end{eqnarray}
Solving this equation, we obtain
\begin{eqnarray}
&&l(\sigma) = 2\left(\frac{l_{0}}{2} -\int_{\sigma}^{\infty} d\sigma
\left(\frac{F_{DBI}(\sigma)}{F_{DBI}(\sigma_{0})}-1\right)^{-\frac{1}{2}}\right).
\label{m32}
\end{eqnarray}
This solution, for a non-zero $\sigma_{0}$  represents a wormhole
with a finite  size. Thus, a new wormhole is formed when the distance separating the
  D3-brane from the  anti-D3-brane  is $l_{0}$. The width of the wormhole at this point is
  $\sigma_{0}$.  Now  we can write
\begin{eqnarray}
\left(\frac{1}{\sqrt{D_{TA}}}TA'(\sigma)\right)'=\frac{1}{\sqrt{D_{TA}}}
\left[\frac{(V(TA)F_{DBI})}{F_{DBI}V(TA)'}(D_{TA}-TA'(\sigma)^{2})\right].
\label{m33}
\end{eqnarray}
Solving this equation, we obtain
\begin{eqnarray}
TA\sim
\sqrt{\frac{\sigma_{0}^{2}}{\sigma_{0}^{2}-\sigma^{2}}}\left(\frac{1}{1+\frac{\pi^{2}N
T^{4}}{6T_{D3}}} \right)\label{m34}.
\end{eqnarray}
So, the  tachyonic states cause   the formation of the wormhole at
$\sigma_{0}=0$. The width of the wormhole increases as the distance between the D3-brane and the anti-D3-brane reduces, 
and finally the D3-brane collides with the anti-D3-brane.

Now using the action given by Eq. (\ref{m30}), we can obtain entropy and mass density along $z$-direction for this  tachyonic system,
 \begin{eqnarray}
 S_{tb}&=& \frac{4T_{D3}^{2}}{\pi T^{4}}\int
d\sigma V(TA(\sigma))\frac{F_{DBI}(\sigma)}{\sqrt{F_{DBI}^{2}(\sigma)-F_{DBI}^{2}(\sigma_{0})}}\sigma^{3}\nonumber \\ && \times
\frac{4}{ \cosh^{4}\alpha}\frac{\sigma_{0}}{(\sigma^{2}-\sigma_{0}^{2})^{3/2}} \label{m35} \\
\frac{dM_{tb}}{dz}&=&\frac{2T_{D3}^{2}}{\pi T^{4}}
 V(TA(\sigma))\frac{F_{DBI}(\sigma)}{F_{DBI}(\sigma_{0})}\sigma^{3}\frac{4 \cosh^{2}\alpha +
1}{ \cosh^{4}\alpha}\frac{\sigma_{0}}{(\sigma^{2}-\sigma_{0}^{2})^{3/2}} \label{m36}.
\end{eqnarray}

We can analyse the effect of tachyonic potential on the number of degrees of freedom of the universe. This can be done  by repeating
the analysis of the   previous section. Thus,  we write again relate    these degrees
of freedom to the entropy of BIon. We also write an expression for the  the mass density
along the transverse direction,
 \begin{eqnarray}
 N_{sur} + N_{bulk} &=& N_{BIon}= N_{brane} + N_{anti-brane} +
N_{wormhole}\nonumber \\&  \simeq &
\frac{4T_{D3}^{2}}{\pi T^{4}}\int
d\sigma V(TA(\sigma))\frac{F_{DBI}(\sigma)}{\sqrt{F_{DBI}^{2}(\sigma)-F_{DBI}^{2}(\sigma_{0})}}
\nonumber \\ && \times \sigma^{3}\frac{4}{ \cosh^{4}\alpha}\frac{\sigma_{0}}{(\sigma^{2}-\sigma_{0}^{2})^{3/2}}
\nonumber \\  N_{sur} - N_{bulk} &\simeq& \int d\sigma
\frac{dM_{BIon}}{dz}\nonumber \\ &=&
\frac{2T_{D3}^{2}}{\pi T^{4}}\int d\sigma
 V(TA(\sigma))\frac{F_{DBI}(\sigma)}{F_{DBI}(\sigma_{0})}\sigma^{3}\frac{4 \cosh^{2}\alpha +
1}{ \cosh^{4}\alpha}\nonumber \\ && \times \frac{\sigma_{0}}{(\sigma^{2}-\sigma_{0}^{2})^{3/2}} \label{m37}.
\end{eqnarray}
Solving these equations simultaneously, we obtain,
\begin{eqnarray}
  N_{sur} &\simeq& \frac{4T_{D3}^{2}}{\pi T^{4}}\int
d\sigma V(TA(\sigma))\frac{F_{DBI}(\sigma)}{\sqrt{F_{DBI}^{2}(\sigma)-F_{DBI}^{2}(\sigma_{0})}} \nonumber \\ &&
\times  \sigma^{3}\frac{4}{ \cosh^{4}\alpha}
\frac{\sigma_{0}}{(\sigma^{2}-\sigma_{0}^{2})^{3/2}}\nonumber \\&&
+ \frac{2T_{D3}^{2}}{\pi T^{4}}\int d\sigma
 V(TA(\sigma))\frac{F_{DBI}(\sigma)}{F_{DBI}(\sigma_{0})}\nonumber \\ && \times \sigma^{3}\frac{4 \cosh^{2}\alpha +
1}{ \cosh^{4}\alpha}\frac{\sigma_{0}}{(\sigma^{2}-\sigma_{0}^{2})^{3/2}}, \nonumber \\   N_{bulk}
&\simeq& \frac{4T_{D3}^{2}}{\pi T^{4}}\int
d\sigma V(TA(\sigma))\frac{F_{DBI}(\sigma)}{\sqrt{F_{DBI}^{2}(\sigma)-F_{DBI}^{2}(\sigma_{0})}}
\nonumber \\ && \times \sigma^{3}\frac{4}{ \cosh^{4}\alpha}\frac{\sigma_{0}}{(\sigma^{2}-\sigma_{0}^{2})^{3/2}}\nonumber \\&&
- \frac{2T_{D3}^{2}}{\pi T^{4}}\int d\sigma
 V(TA(\sigma))\frac{F_{DBI}(\sigma)}{F_{DBI}(\sigma_{0})}\nonumber \\ && \times \sigma^{3}\frac{4 \cosh^{2}\alpha +
1}{ \cosh^{4}\alpha}\frac{\sigma_{0}}{(\sigma^{2}-\sigma_{0}^{2})^{3/2}} \label{m38}.
\end{eqnarray}

This equation demonstrates  that as the D3-brane approaches the anti-D3-brane,
tachyonic potential increases and tends to infinity. Thus the
 number of degrees of freedom becomes large as one trends to
the  Big Rip  singularity. Thus, for this system,  the Hubble parameter
for flat universe can be written as
\begin{eqnarray}
 H_{flat,ac} \simeq  \left(\frac{1}{V(TA)}\right)^{1/2}\left(\frac{72\pi k^{8}N^{14}
T_{F1}^{16}}{T_{D3}^{16}\sigma_{0}^{10}} T^{14} + \frac{8\pi
k^{4}N^{12} T_{F1}^{14}}{T_{D3}^{14}\sigma_{0}^{6}}
T^{12}\right)\label{m39}.&&
\end{eqnarray}
It may be noted that the
Hubble parameter depends on tachyonic potential
between the D3-brane and the anti-D3-brane.
As the   tachyonic potential increases, Hubble parameter reduces to a very small value.

Finally, using the Friedmann equation for the flat Friedmann-Robertson-Waker
universe, we can write the   energy density for this system as
\begin{eqnarray}
  \rho_{flat,ac} &=&  \frac{3}{8\pi l_{P}^{2}}H^{2}
  \nonumber \\ &\simeq&
\frac{3}{8\pi l_{P}^{2}}
 \left(\frac{1}{V(TA)} \right) \left(\frac{4900\pi k^{16}N^{24}
T_{F1}^{32}}{T_{D3}^{32}\sigma_{0}^{120}} T^{28} + \frac{16\pi
k^{8}N^{24} T_{F1}^{28}}{T_{D3}^{28}\sigma_{0}^{12}}T^{24} \right. \nonumber \\ && \left. +
\frac{496\pi k^{12}N^{26}
T_{F1}^{30}}{T_{D3}^{30}\sigma_{0}^{16}}T^{26}
\right).
\label{m40}
\end{eqnarray}
The energy density decreases with increasing
tachyonic states,  and it reduces  to zero at $TA=\infty$.
This is because that D3-brane moves towards the anti-D3-brane,
 and this creates  tachyon states, which in turn
increase the  radius of universe. This leads to acceleration of the universe,
and it also decreases  the energy density of the universe.

The  properties of open and closed universes can now be obtained from
Eq.   (\ref{m17}),
and  $r_{A} = 1/ \sqrt{H^{2} + \frac{\bar{k}}{a^{2}}}$. So, the  Hubble parameter
for non-flat universe during this stage of evolution can be expressed as
\begin{eqnarray}
  H_{o/c,ac} &\simeq& \sqrt{\left(\frac{1}{V(TA)}\right)}\nonumber \\ && \times \sqrt{\left(\frac{72\pi k^{8}N^{14}
T_{F1}^{16}}{T_{D3}^{16}\sigma_{0}^{10}} T^{14} + \frac{8\pi
k^{4}N^{12} T_{F1}^{14}}{T_{D3}^{14}\sigma_{0}^{6}}
T^{12}\right)^{2}-\bar{k}/a^{2}}\label{mm19}.
\end{eqnarray}
Now we can define  $\mathcal{T}_t$ as
\begin{eqnarray}
 \mathcal{T}_t &=& \left(\frac{72\pi k^{8}N^{14}
T_{F1}^{16}}{T_{D3}^{16}\sigma_{0}^{10}} T^{14} + \frac{8\pi
k^{4}N^{12} T_{F1}^{14}}{T_{D3}^{14}\sigma_{0}^{6}}
T^{12}\right)^{2}.
\end{eqnarray}
Using this definition of $\mathcal{T}_t$, we can  write  the scale factor for open   universe as
\begin{eqnarray}
a_{o,ac}(t) \simeq \exp-\int dt  [ \mathcal{T}_t + ln(t) ],&& \label{mmm19}
\end{eqnarray}
and the scale factor for  closed universe as
\begin{eqnarray}
 a_{c,ac}(t) \simeq \exp -i\int dt  [ \mathcal{T}_t + ln(t) + \frac{\pi}{2}  ] .&& \label{mmmm19}
\end{eqnarray}
 These scale factors depend on both the tachyonic potential and the temperature.
 Thus, by  increasing tachyons,
 the scale factor of open universe increases and tends to infinity at $TA=\infty$,
 however, the scale factor of closed universe oscillates at this stage.

 The energy density of the open and closed universes can be written as
 \begin{eqnarray}
  \rho_{o/c,ac} &=& \frac{3}{8\pi l_{P}^{2}}(H_{o/c,ac}^{2}+k/a^{2})  \nonumber \\&\simeq&
   \frac{3}{8\pi l_{P}^{2}}H_{flat,ac}^{2} \nonumber \\ &\simeq&
   \frac{3}{8\pi l_{P}^{2}} \left(\frac{1}{V(TA)}\right)\left(\frac{72\pi k^{8}N^{14}
T_{F1}^{16}}{T_{D3}^{16}\sigma_{0}^{10}} T^{14} + \frac{8\pi
k^{4}N^{12} T_{F1}^{14}}{T_{D3}^{14}\sigma_{0}^{6}}
T^{12}\right)^{2} \nonumber \\ &=&
\rho_{flat,ac}.
\label{mmmmm19}
\end{eqnarray}
It may be noted that again the energy density for the open universe is equal to the energy density of the closed
universe.  This is because this energy density depends on the
  tachyonic potential, and not on the specific type of the universe.
  Thus, we have been able to analyse the state of the universe just before a D3-brane collides with an anti-D3-brane.

\section{Conclusion} \label{sum}
In this paper, we used    a   configuration of a D3-brane and an anti-D3-brane connected by a wormhole  for analysing
the holographic inflationary cosmology. This BIonic solution was obtained from
    fundamental black strings. The flow of degrees of freedom into the D3-brane caused inflation.
    This flow occurred due to the wormhole connecting the D3-brane with the anti-D3-brane, and so the inflation ended when this wormhole was annihilated.
   We also pointed out that it is possible for a D3-brane to collide with an anti-D3-brane.
    Such a collision will occur when the distance between
 between a D3-brane and an anti-D3-brane is  reduced below a critical value. However, the reduction of the distance between
 a D3-brane and an anti-D3-brane will lead to the formalism of
  tachyonic states.  A new wormhole  will form   due to the presence of these
 tachyonic states, and this will cause the  late time acceleration of the universe,  before such an collision.
 It may be noted that  the holographic cosmology  \cite{1}-\cite{2}  has
already been used for analyzing    brane world models  \cite{f1}, we have generalized such an analysis to
a BIonic solution.   Even though  it had been proposed that   holographic cosmology can be analysed using the BIon \cite{f8},
in this paper, we have analysed the detailed consiquences of such a model. We have obtained explicit expression for the Hubble parameter
and energy density of the universe for such a model.

 It may be noted that the finite temperature effects for
 non-extremal self-dual string  solutions and wormhole solutions
 interpolating between stacks of M5-branes and anti-M5-branes have also been studied, and such
   solutions define a BIon solution in M-theory  \cite{w}-\cite{w1}.
 This analysis has been done     using the blackfold approach, and
  the self-dual string solitonic solutions
have been analysed as a three-funnel solution of an effective five-brane theory.
 It would be interesting to study holographic cosmology using  this BIonic solution.
 Furthermore,      solution to the non-abelian theory of coincident D-strings has been analysed using
 noncommutative geometry
 \cite{ds}. This was done by     using funnels to  the non-abelian D-string expanding out into an orthogonal D3-brane.
 It was also demonstrated that this  configurations is dual to the BIonic solutions in the abelian worldvolume theory of the D3-brane.
 In this paper, we have analysed the holographic cosmology using the BIonic solutions, so it would be interesting to analyse the
  holographic cosmology using this theory which is dual to a BIonic solution.
  It is also possible to include   the effects of a nonzero $NS-NS$ two-form field in this dual theory \cite{sd}.
 A tilted BIon is obtained by the inclusion of such a field, and   the core of this tilted BIon expands out to a single D3-brane (at an angle to the D1-brane core).
 It has been demonstrated that the  properties of this system are consistent with  that of
  an abelian D3-brane in a background worldvolume magnetic field. It would be interesting to analyse the  holographic cosmology using such a system.

 The  strings have an extended structure, and the extended structure of strings
 is expected to act as a minimum measurable length scale in spacetime
 \cite{s1}. This is because the    smallest probe avilable  in string theory
 is the fundamental string. So, it is not possible to probe spacetime below string length scale.
 However, the existence of a minimum length scale in spacetime leads to the
 generalized uncertainty principle \cite{s2}-\cite{2s}.
 In fact, it has been demonstrated that the generalized uncertainty principle can be used to analyse the
 corrections to the $AdS/CFT$ correspondence  beyond the supergravity approximation \cite{faiz}.
  It may be noted that the corrections to Friedmann equations coming from the
 generalized uncertainty principle has been obtained
 \cite{q7}-\cite{q9}.  It would be interesting to analyse the effect of extended structure of strings on the holographic cosmology. It is expected that it will deform
 the BIonic holographic cosmology.
 The effect of generalized uncertainty principle on the usual holographic cosmology has already been analysed \cite{gene}. Thus, it would be interesting
 to repeat  that analyses  for the BIonic holographic cosmology. It may be noted that even thought the holographic cosmology has been studied for
 brane world models \cite{b0}, cosmological models in scalar-tensor gravity \cite{b1},
and cosmological models in $f(R)$ gravity \cite{b2}, the deformation of these systems by generalized uncertainty has not been analysed. However, it is possible
to study such a generalization, and it would be interesting to analyse the deformation of these cosmological models using the generalized uncertainty principle.
It may be noted that  cosmological and astrophysical consiquences of extended theories
of gravity have been studied in both
metric and palatini formalism \cite{pala}. It would be interesting to relate the work done
in this paper to similar cosmological and astrophysical observations.
It may be noted that $\Lambda$CDM model and
other  observational tests have been used to  constrain the current cosmic acceleration \cite{cosmi}.
It would be important to use the    $\Lambda$CDM and these other observational tests to constraint
the parameters of  holographic cosmology.

\section*{Acknowledgments}
\noindent
Authors wish to thank  Professor Rahaman and Professor Pradhan for helpful discussions. The work of Alireza Sepehri has been supported financially by Research Institute for Astronomy  Astrophysics of Maragha (RIAAM) under research project No.1/4717-96.


\begin{thebibliography}{99}
\bibitem{a}E. S. Fradkin and A. A. Tseytlin,  Phys. Lett. B 163, 123 (1985)
\bibitem{a1}A. Abouelsaood, C. G. Callan,   C. R. Nappi  and
S. A. Yost,   Nucl. Phys. B 280, 599  (1987)

\bibitem{b}C. G. Callan and J. M. Maldacena,
Nucl. Phys. B 513, 198 (1998)
\bibitem{b4}R. Emparan,
Phys. Lett. B 423, 71 (1998)

\bibitem{c}G. W. Gibbons,   Nucl. Phys. B 514, 603
(1998)


\bibitem{d}J. M. Maldacena,
Adv. Theor. Math. Phys. 2, 231 (1998)

\bibitem{e}S. S. Gubser, I. R.
Klebanov  and A. M. Polyakov, Phys. Lett. B 428,  105 (1998)

\bibitem{f} E. Witten,
  Adv. Theor. Math. Phys. 2,  253 (1998)
\bibitem{g}J. McGreevy, L. Susskind  and N. Toumbas,  JHEP 06,  008 (2000)

 \bibitem{h}M. T. Grisaru, R. C. Myers  and O. Tafjord,   JHEP 08,
040 (2000)

 \bibitem{i}S. J. Rey and J. T. Yee,  Eur. Phys. J. C 22, 379   (2001)

 \bibitem{j}J. M. Maldacena,   Phys. Rev. Lett. 80,
4859  (1998)

\bibitem{4}J. Jacobson, Phys. Rev. Lett. 75, 1260 (1995)
\bibitem{8}E. P. Verlinde, JHEP 1104, 029 (2011)

\bibitem{1}T. Padmanabhan, Class. Quant. Grav. 21, 4485 (2004)
\bibitem{2}T. Padmanabhan, arXiv:1206.4916 (2012)
\bibitem{9}K. Yang. Y. X. Liu and Y. Q. Wang, Phys. Rev. D 86, 104013 (2012)
\bibitem{b0}Y. Ling and J. P. Wu, JCAP, 1008, 017 (2010)

 \bibitem{b1}M. Akbar and R.  G. Cai, Phys. Lett. B 635, 7 (2006)

\bibitem{b2}S. Capozziello, V. F. Cardone, and A. Troisi, Phys. Rev. D 71, 043503
(2005)


\bibitem{f1}Y.  Ling and W. J.  Pan, Phys. Rev. D 88, 043518 (2013)

\bibitem{f2}A.  Sheykhi, Phys. Rev. D 87, 061501 (2013)

\bibitem{f4}M.  Eune and  W.  Kim, Phys. Rev. D 88, 067303 (2013)

\bibitem{f6}E.  C. Young  and D. Lee, JHEP 04, 125 (2014)


\bibitem{f8}A. Sepehri, F. Rahaman, A. Pradhan and  I. H. Sardar, Phys. Lett. B 741, 92 (2014)

\bibitem{f9}G. Grignani, T.  Harmark, A.  Marini, N.  A. Obers and  M.  Orselli, JHEP 1106, 058 (2011)

\bibitem{f10}G. Grignani, T.  Harmark, A.  Marini, N.  A. Obers and  M.  Orselli, Nucl. Phys. B 851, 462 (2011)

\bibitem{ff10} M. R. Setare and A. Sepehri, JHEP 03, 079 (2015)


 \bibitem{m1}  G. Grignani, T.  Harmark, A.  Marini and M.  Orselli, JHEP 1403, 114 (2014)

 \bibitem{n1}M. R. Garousi, JHEP 0501,  029 (2005)
 \bibitem{n2}A.  Dhar and P.  Nag, JHEP 0801, 055 (2008)
 \bibitem{n4}A.  Dhar and
P.  Nag, Phys. Rev. D 78, 066021 (2008)
 \bibitem{n5}M. R. Setare, A. Sepehri and  V. Kamali, Phys. Lett. B 735, 84 (2014)
 \bibitem{m2}A. Sen, Phys. Rev. D 68, 066008 (2003)

  \bibitem{w} V.  Niarchos and  K.  Siampos, JHEP 1206, 175 (2012)
 \bibitem{w1}V.  Niarchos and  K.  Siampos, JHEP 1207, 134 (2012)

\bibitem{ds}  N. R. Constable, R. C. Myers and O. Tafjord, Phys. Rev. D 61, 106009 (2000)
\bibitem{sd} J. L. Karczmarek and C. G. Callan,  JHEP 0205, 038 (2002)

   \bibitem{s1}D. Amati, M. Ciafaloni and G. Veneziano, Phys. Lett. B 216, 41 (1989)

   \bibitem{s2} A. Kempf, G. Mangano, and R. B. Mann,  Phys. Rev. D   52,  1108 (1995)
   \bibitem{127q} S. Das, and E. C. Vagenas,  Phys. Rev. Lett.   101,  221301 (2008)
\bibitem{120000a1} M. Liu, Y. Gui and H. Liu,  Phys. Rev. D78, 124003 (2008)
\bibitem{120000a2}R. Garattini and M. Faizal,  Nucl. Phys. B905, 313 (2016)
\bibitem{12greene}R. Easther, B. R. Greene, W. H. Kinney and G. Shiu, Phys.  Rev.  D {  67}, 063508 (2003)
\bibitem{fgdghp} M. Faizal and  M. M. Khalil, Int. J. Mod. Phys. A 30, 1550144 (2015)
\bibitem{12g}   L. J. Garay, Int. J. Mod. Phys. A 10, 145  (1995)
\bibitem{12g1}M. Maggiore, Phys. Lett. B 304,   65 (1993)
 \bibitem{2s}A. F. Ali, S. Das and E. C. Vagenas, Phys. Rev. D   84 , 044013 (2011)

 \bibitem{faiz}  M. Faizal, A. F. Ali and  A. Nassar,  Int. J. Mod. Phys. A 30, 1550183 (2015)

 \bibitem{q7} P.  Bargueno and  E.  C. Vagenas,  Phys. Lett. B 742,  15  (2015)
   \bibitem{q9} A.  Awad and A. F.  Ali,  JHEP 1406,  093  (2014)

   \bibitem{gene}A. F. Ali, Phys. Lett. B 732, 335 (2014)
   \bibitem{pala}  S. Capozziello and M. Francaviglia,  	Gen. Rel. Grav. 40, 420 (2008)
\bibitem{cosmi} K. Bamba, S. Capozziello, S.  Nojiri and S. D. Odintsov,
  Astrophys.Space Sci.  342, 155 (2012)
 \end{thebibliography}
 \end{document}